\documentclass[%
 reprint,
superscriptaddress,
 amsmath,amssymb,
 aps,
 prl,
]{revtex4-2}

\usepackage{amsmath}
\allowdisplaybreaks
\usepackage{graphicx}
\usepackage{amssymb, amsfonts, amsmath,framed}
\usepackage{bm}
\usepackage{nicefrac}
\usepackage[colorlinks=true]{hyperref}
\usepackage[dvipsnames]{xcolor}
\hypersetup{
    colorlinks=true,
    allcolors=Violet,
}
\usepackage[shortlabels]{enumitem}

\newcommand{\dd}{\mathrm{d}}

\newcommand{\bx}{\boldsymbol{x}}

\newcommand{\bz}{\boldsymbol{z}}

\newcommand{\bu}{\boldsymbol{u}}

\newcommand{\bF}{\boldsymbol{f}}

\newcommand{\bnabla}{\boldsymbol{\nabla}}
\newcommand{\bcdot}{\boldsymbol{\cdot}}

\newcommand{\bB}{\boldsymbol{B}}
\newcommand{\bE}{\boldsymbol{E}}

\newcommand{\bb}{\boldsymbol{b}}

\newcommand{\p}{\partial}

\newcommand{\blue}[1]{\textcolor{black}{#1}}

\begin{document}

\title{Forced 3D reconnection in an exponentially separating magnetic field}

\author{David N. Hosking}
\email{dnh26@cam.ac.uk}
\affiliation{Princeton Center for Theoretical Science, Princeton, NJ 08540, USA}
\affiliation{Gonville \& Caius College, Trinity Street, Cambridge, CB2 1TA, UK}
\author{Ian G. Abel}
\affiliation{Institute for Research in Electronics and Applied Physics, University of Maryland, College Park, MD 20742, USA}
\author{Steven C. Cowley}
\affiliation{Princeton Plasma Physics Laboratory, Princeton, NJ 08540, USA}

\date{\today}

\begin{abstract}
We present a solvable scenario for 3D reconnection in a sheared magnetic field. We consider a localized external force that is applied slowly \blue{to a flux tube} and then maintained, generating an Alfvénic perturbation that spreads along the field lines. Separation of the sheared field lines \blue{reduces} the scale of the perturbation across the field, \blue{enhancing magnetic diffusion}. For a fusion-motivated equilibrium with exponential field-line separation, we find a reconnection timescale proportional to $\mathcal{S}/\ln \mathcal{S}$ under magnetohydrodynamics (MHD) and to $\mathcal{S}^{1/3}$ for semicollisional electron-only reconnection, where $\mathcal{S}$ is the Lundquist number of the perturbed flux tube. We generalize these results to arbitrary magnetic geometries, showing that the \blue{semicollisional case} is geometry independent. Interestingly, we find that slower field-line separation yields an increased reconnection rate in MHD.
\end{abstract}

\maketitle
\textit{Introduction.\quad}
\blue{Magnetic reconnection---the topological reconfiguration of a magnetic field in a conducting medium---is a fundamental process of key importance in both laboratory and astrophysical plasmas. In fusion devices, reconnection can cause sawtooth crashes~\cite{Kadomtsev75, Liu24} and act both as trigger~\cite{Cathey20} and saturation mechanism~\cite{Ham18} for edge-localized modes (ELMs). In astrophysical and space plasmas, reconnection mediates and drives turbulence~\cite{LoureiroBoldyrev17, Schekochihin20}, accelerates particles~\cite{SironiSpitkovsky14} and enables solar flares~\cite{Drake25}. While reconnection in 2D is enabled by the tearing instability~\cite{FKR63,Coppi76} (particularly in its nonlinear~\cite{Uzdensky10} and/or kinetic~\cite{DrakeLee77} regimes), it has been suggested that alternative 3D mechanisms may facilitiate faster reconnection~\cite{Boozer18, Boozer19, Lazarian20}. In this Letter, we present a solvable case of 3D reconnection occurring via a 3D analogue of resistive tearing. The process we consider occurs when a force is applied locally to a sheared flux tube and maintained. This induces a resistively damped Alfv\'{e}n-wave packet that thins as it propagates along separating field lines. Because the packet is associated with a local change in the direction of the magnetic field, its resistive diffusion modifies the connectivity of the global field.}

\begin{figure*}
    \centering
    \includegraphics[width=0.93\textwidth]{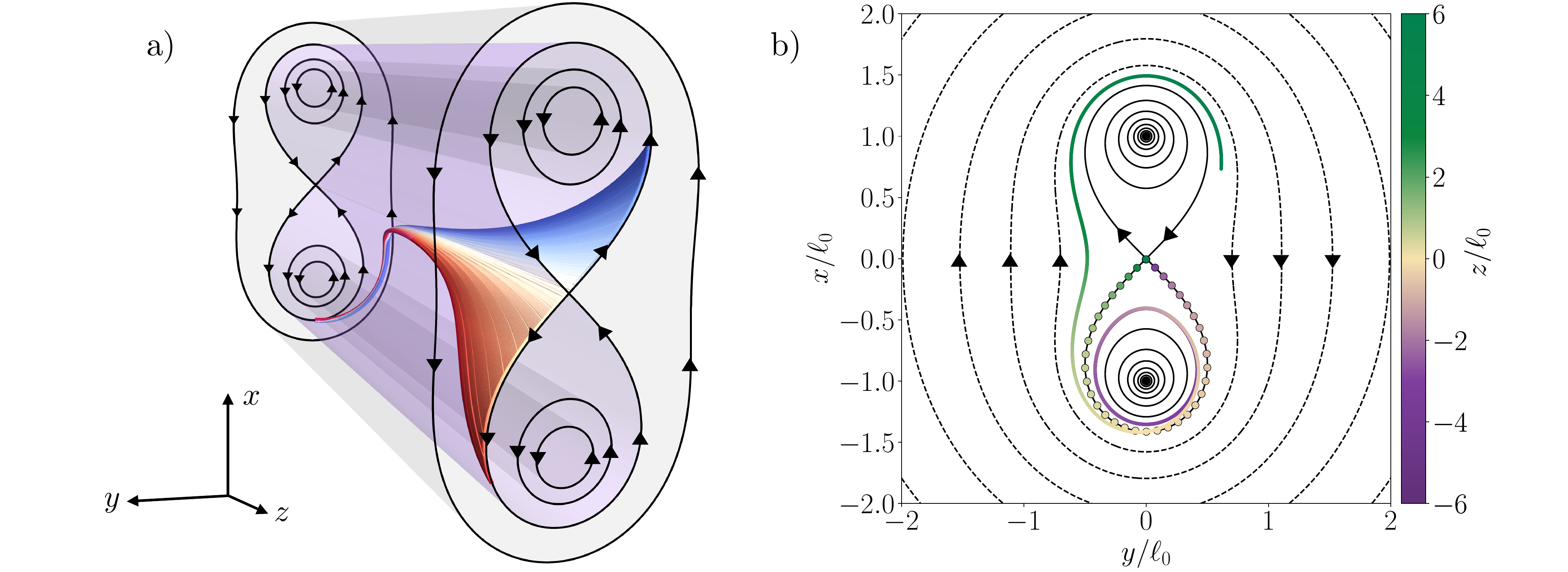}
    \caption{Panel (a): The magnetic geometry considered in this study [Eq.~\eqref{twocurrent}]. The $\alpha$ surface along which we consider displacing a field line is shown with $\psi$ increasing from red to blue. Arrows show the projection of the magnetic-field direction onto the $xy$-plane.
    Panel (b): Projection onto the $xy$-plane of the field line through $(-\sqrt{2}a,0,0)$ at (\textit{i}) $\tilde{t}=0$ (dots with black outlines) and (\textit{ii}) $\tilde{t}\gg \tau_{\mathrm{rec}}$ \blue{(purple-green line)}. Note that the orientation of the $y$-axis in panel~(b) is opposite to that in panel~(a).}
    \label{fig:illustration}
\end{figure*}

\blue{Although we generalize to arbitrary equilibria at the end of the Letter, we develop and illustrate our results for} a particular fusion-motivated equilibrium [Eq.~\eqref{twocurrent}] with field lines that separate exponentially with the distance~$\ell$ along them [Fig.~\ref{fig:illustration}(a)]. It has been suggested, and is perhaps intuitively compelling, that the rate of reconnection of exponentially separating field lines is fast, i.e., at most logarithmic in the Lundquist number~$\mathcal{S}$ (the ratio of the ideal to nominal resistive timescales)~\cite{Boozer18, Boozer19}.  It seems intuitive that this would be the case for the scenario described above: a diffusing Alfv\'{e}n wave would be sheared to resistive scales in~$\ln \mathcal{S}$ Alfv\'{e}n timescales. However, we find that the reconnection timescale remains proportional to a finite power of~$\mathcal{S}$ (see abstract) \blue{because the field-line separation arrests propagation of the wave. Indeed, we find \textit{faster} reconnection for field lines that separate more slowly (provided that they separate faster than $\ell^{1/2}$}). 

\textit{Clebsch coordinates.\quad}We represent the equilibrium magnetic field $\boldsymbol{B}_0$ with Clebsch coordinates $\psi$ and $\alpha$, i.e., $\bm{B}_0 = \bnabla\psi\times\bnabla \alpha$.
We take $\psi$ to have units of magnetic flux and $\alpha$ to be dimensionless. The Clebsch coordinates are constant along unperturbed field lines and can therefore be used to label them; the arc length $\ell$ along the \blue{field} line completes the coordinate set. \blue{We take the applied force $\bF$ to be localized to, and directed along, a particular $\alpha$-surface (i.e., $\bF \propto \bB_0\times \bnabla \alpha$). The distance perpendicular to~${\bB}_0$ between two field lines $(\psi, \alpha)$ and $(\psi + d\psi, \alpha)$ in an $\alpha$ surface is}
\begin{equation}
\dd l_\perp =  \frac{|\bnabla\alpha |}{B_0}\dd\psi,
\label{distance}
\end{equation}so the rate of separation of the field lines is encoded in the dependence of $|\bnabla \alpha |$ on $\ell$. This dependence determines the reconnection rate in our theory. \blue{(Note that the coordinate $\alpha$ is not uniquely defined for given~$\bB_0$. Different definitions correspond to different applied forces, which may be associated with different reconnection rates.)}

\textit{Model equilibrium.\quad}We shall consider as a model magnetic geometry the field of two wires, located at $x=\pm a$ and $y=0$, each carrying current $I_0$ in the $z$ direction. We embed this configuration in a uniform external field $B_{\mathrm{ext}}\hat{\bz}$, so that, with $r^2 = x^2 + y^2$ and $c$ the speed of light,
\begin{equation}
 \bm{B}_0 = B_{\mathrm{ext}}{\hat{\bz}} +  {\hat{\bz}}\times \frac{\bnabla\psi}{a},\,\,\, \psi = \frac{a I_0}{c} \ln\left[(r^2 + a^2)^2 - 4a^2x^2\right].\label{twocurrent}\end{equation}The equilibrium~\eqref{twocurrent} is visualized in Fig.~\ref{fig:illustration}(a). Toroidicity notwithstanding, each lobe resembles the magnetic field of a tokamak in a divertor configuration (with the lobes taken together, it resembles a ``doublet''~\cite{Jensen75}). \blue{We expect the results of our Letter to be applicable to reconnection in such a device, forced, for example, by an ELM-control coil (whose effect on the topology of edge fields and transport may be crucial for confinement~\cite{Evans08, Wade15, Ryan24}).}

\blue{We consider an odd-parity force that acts in the vicinity of the point $(-\sqrt{2}a, 0,0)$ on the separatrix, inducing reconnection of field lines passing through this region}. As we explain in Section 1 of the End Matter, we may define~$\alpha$ such that
\begin{equation}
|\bnabla\alpha|^2 = \frac{\ell_0^2}{2a^4}\cosh{\left(\frac{2z}{\ell_0}\right)} +\frac{2a^2}{\ell_0^2} + \frac{36(z/\ell_0)^2}{\cosh{(2z/\ell_0)}}\label{gradalpha}
\end{equation}on this field line, where $\ell(z)$ is the solution of
\begin{equation}
    \frac{\dd \ell}{\dd z} = \frac{B_0}{B_{\mathrm{ext}}} = \sqrt{1 + \frac{2a^2}{\ell_0^2\cosh{(2z/\ell_0)}}},\label{dldz}
\end{equation}$\ell_0 = a^2 B_{\mathrm{ext}}/4I_0$ is the magnetic-shear length (we take $\ell_0=a$ in all numerical results presented in this Letter) and $B_{0}(\ell)=|\bB_0|$. \blue{This choice corresponds to a force in the $\hat{\bx}$ direction.} According to Eqs.~\eqref{distance} and~\eqref{gradalpha}, field lines separate exponentially with $\ell$ as ${\ell \to\infty}$ \blue{[see Fig.~\ref{fig:illustration}(a)]}.

\textit{Dynamical model.\quad}We take the equilibrium magnetic field to be strong, and therefore consider a perturbation that is extended along the field line. We consider two dynamical regimes: the \textit{MHD case}, in which the perpendicular scale of the perturbation is much larger than the ion gyroradius $\rho_i$, and the \textit{kinetic case}, in which it is much smaller than $\rho_i$. \blue{We employ the equations of low-$\beta$ reduced MHD~\cite{Strauss76} to describe the MHD case and the semi-collisional limit of the Kinetic Reduced Electron Heating Model (KREHM) of Ref.~\cite{ZoccoSchekochihin11} to describe the kinetic case (see Section~1 of the Supplementary Information)}. The latter regime is interesting \textit{a priori} because the parallel group velocity of \textit{kinetic} Alfv\'{e}n waves is an increasing function of their perpendicular wavenumber~\cite{Schekochihin09}: the equilibrium shear causes them to accelerate and thus reach diffusive scales more quickly~\footnote{We note that the semi-collisional limit (collisionless ions but collisional electrons) means that these waves are immune to Landau damping.}.

We assume an eikonal form for the velocity streamfunction~$\Phi$ and perturbed magnetic-flux function~$\Psi$, viz., $\Phi(\psi, \alpha, \ell, t) = \bar{\Phi}(\ell, t)e^{in\alpha}$ and $\Psi(\psi, \alpha, \ell, t) = \bar{\Psi}(\ell, t)e^{in\alpha}$, where $n \gg 1$ (cf. Refs.~\cite{Roberts65,Connor78,Conner85,DrakeAntonsen85}). As we explain in Section~1 of the Supplementary Information, this yields
\begin{equation}
    \left(\begin{cases} 
        |\tilde{\bnabla} \alpha|^2 & (\text{MHD}) \\
        1 & (\text{kinetic})
    \end{cases}\right)\times\frac{1}{\tilde{B}_0(\tilde{\ell})^2}\frac{\partial \tilde{u}}{\partial \tilde{t}} 
    = \frac{\p \tilde{J}_{\parallel}}{\p \tilde{\ell}}+\tilde{F}(\tilde{\ell}, \tilde{t}),\label{vorticity}
\end{equation}and
\begin{equation}
    \frac{1}{|\tilde{\bnabla} \alpha|^2}\frac{\p \tilde{J}_\parallel}{\p \tilde{t}}=\frac{1}{\tilde{B}_0(\tilde{\ell})}\frac{\p [\tilde{B}_0(\tilde{\ell})\tilde{u}]}{\p \tilde{\ell}}-\frac{\tilde{J}_{\parallel}}{\mathcal{S}}.\label{induction}
\end{equation}

\begin{figure}
    \centering
    \includegraphics[width=\columnwidth]{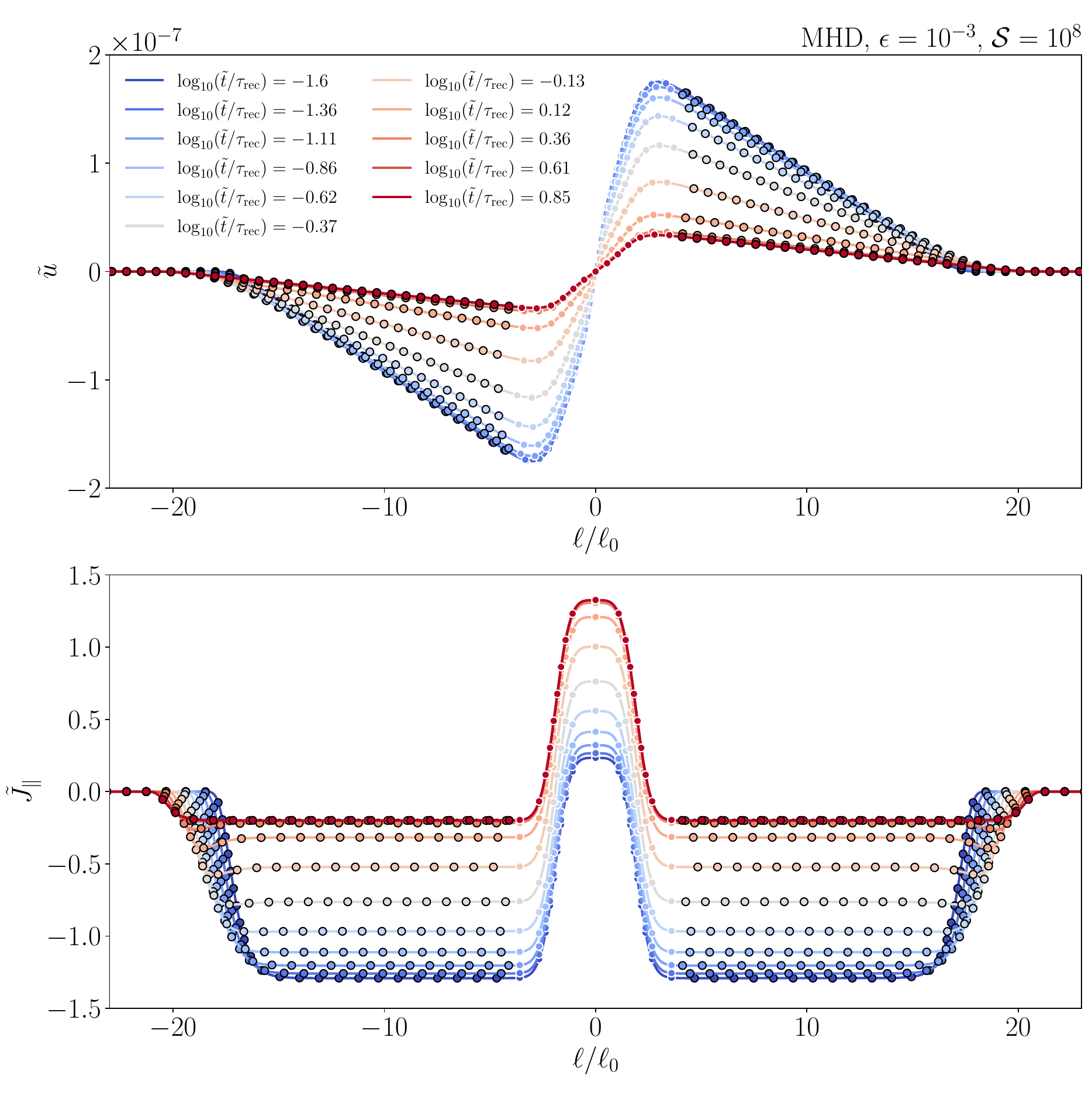}
    \caption{Evolution of $\tilde{u}$ and $\tilde{J}_{\parallel}$ for $\tilde{t}\sim \tau_{\mathrm{rec}}$ in the MHD case. Solid lines show numerical solutions of Eqs.~\eqref{vorticity} and~\eqref{induction} obtained with a finite-difference scheme. Dots show the theoretical prediction derived in the main text: dots with black outlines correspond to the outer solution [Eqs.~\eqref{usolnouterMHD} and~\eqref{JsolnouterMHD}  (corrected for the resistive diffusion of the inner solution, as described in Section 4 of the End Matter)], dots with white outlines to the inner solution [Eqs.~\eqref{JsolninnerMHD} and~\eqref{usolninnerMHD}].}
    \label{fig:solutions}
\end{figure}

\begin{figure*}
    \centering
    \includegraphics[width=\textwidth]{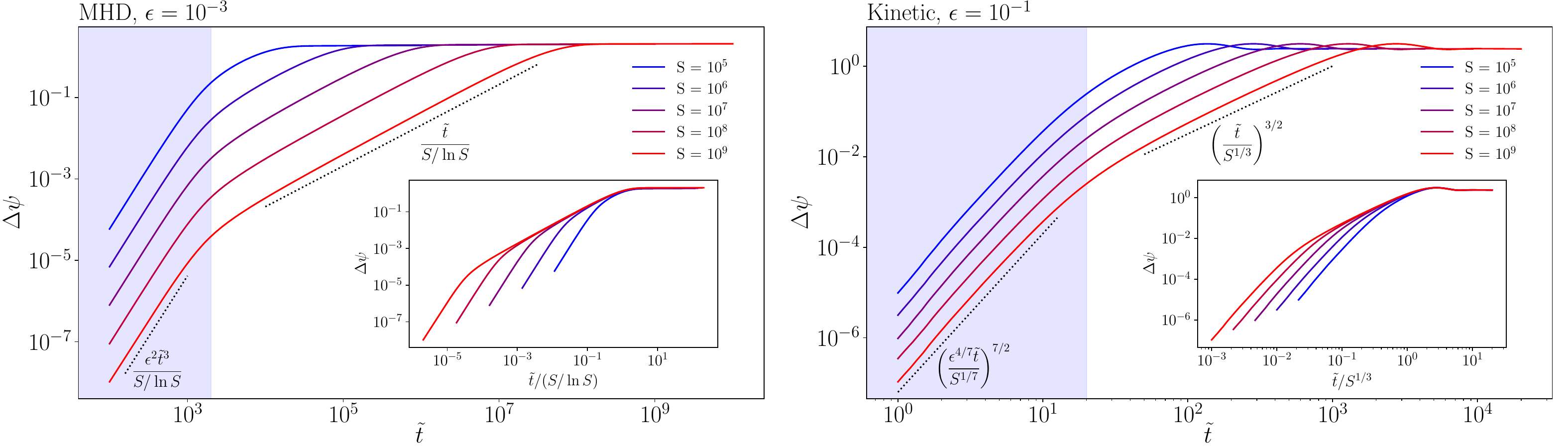}
    \caption{Evolution of $\Delta \psi$ [Eq.~\eqref{Deltapsi}] as a function of time in the MHD (left) and kinetic (right) cases in the numerical solution of Eqs.~\eqref{vorticity} and~\eqref{induction}. At $\tilde{t}\gg 1/\epsilon$, the evolution is as predicted by Eqs.~\eqref{Deltapsi_MHD} and~\eqref{Deltapsi_kinetic} (see insets, which show time rescaled by the appropriate function of $\mathcal{S}$). Blue regions indicate $\tilde{t}\lesssim \epsilon^{-1}$, when the force $\tilde{F}$ \blue{is increasing in size}. The power laws for $\Delta \psi$ in these regions are derived in Section 2 of the Supplementary Information.}
    \label{fig:Deltapsi}
\end{figure*}

The variables that appear in Eqs.~\eqref{vorticity} and~\eqref{induction} are as follows. The velocity, directed parallel to the~$\alpha$ surface, is ${\delta \bu_{\perp} = \bb_0 \times \bnabla \Phi= u(\psi, \alpha, \ell, t) e^{in\alpha} \bb_0 \times \bnabla \alpha}$; ${\delta \bB_{\perp} = \bb_0\times \bnabla \Psi= \delta B_{\perp}(\psi, \alpha, \ell, t) B_0(\ell) e^{in\alpha} \bb_0 \times \bnabla \alpha}$ is the magnetic-field perturbation in the same surface; $\bb_0(\ell)=\bB_0/B_0$; and $J_{\parallel}=|\bnabla \alpha|^2 \delta B_{\perp}$ is the perturbed parallel-current \blue{density}. Tildes signify normalization to the dimensionally appropriate combination of the shear length $\ell_0$ and the ideal timescale $ \tau_A$. The latter is the shear-length crossing time of, in the MHD case, an Alfv\'{e}n wave  (i.e., $\tau_A= \ell_0/v_{A,\mathrm{ext}}$, with $v_{A,\mathrm{ext}}$ the Alfv\'{e}n speed associated with $B_{\mathrm{ext}}$) or, in the kinetic case, a kinetic Alfv\'{e}n wave (i.e., $\tau_A= \sqrt{2}\ell_0^2/n \rho_i v_{A,\mathrm{ext}}$). Similarly normalized, $\tilde{F}(\tilde{\ell}, \tilde{t})$ represents a force with the same eikonal dependence on~$\alpha$ as $\Phi$ and $\Psi$, which pushes the field line along the $\alpha$ surface. We take $\tilde{F}$ to have odd parity in $\tilde{\ell}$ and to turn on smoothly on the (normalized) timescale $\epsilon^{-1}  \gg 1$, reaching a constant value at $\tilde{t}\gg \epsilon^{-1}$. In the numerical results shown in Figs.~\ref{fig:illustration}(b),~\ref{fig:solutions} and~\ref{fig:Deltapsi}, we choose $\tilde{B}_0(\tilde{\ell})^2\tilde{F}(\tilde{\ell}, \tilde{t}) = \tilde{\ell}^3 e^{-\tilde{\ell}^2}(1-e^{-\epsilon^2 \tilde{t}^2})$, absorbing the amplitude of $\tilde{F}$ into the definitions of $\tilde{u}$, $\delta \tilde{B}_{\perp}$ and $\tilde{J}_{\parallel}$. \blue{We define the Lundquist number for the perturbed flux tube to be $\mathcal{\mathcal{S}} = \tau_A/\tau_\eta$, where $\tau_\eta = \ell_0^2 / n^2 \eta$ is the resistive timescale at the perpendicular scale $\ell_0/n$. Thus,}
\begin{equation}
\mathcal{\mathcal{S}} = \begin{cases} 
        S /n^2 \sim 10^4 (100/n)^2 & (\text{MHD}) \\
        \rho_{\star} S /\sqrt{2}n \sim 10^4 (100/n) & (\text{kinetic})
    \end{cases}
\label{resistivegrow}\end{equation}where $S=v_{A,\mathrm{ext}}\ell_0/\eta$ is the Lundquist number of the global equilibrium, $\eta$~is the magnetic diffusivity and ${\rho_{\star}\equiv \rho_i/\ell_0}$ is the normalized ion gyroradius. The numerical values quoted in Eq.~\eqref{resistivegrow} correspond to \blue{$S\sim 10^8$} and $\rho_{\star}\sim 10^{-2}$, which are typical for a tokamak.

Eqs.~\eqref{vorticity} and~\eqref{induction} are linear in~$\tilde{u}$ and~$\tilde{J}_{\parallel}$ even though we have not required their amplitudes to be small compared with their perpendicular wavelengths. In this sense, the solutions we derive are valid nonlinearly~\footnote{We require amplitudes to be small compared with the equilibrium scales in order for Eq.~\eqref{gradalpha} to be valid for the perturbed magnetic field.}.

\textit{Reconnected flux.\quad} \blue{In the absence of the resistive final term, Eq.~\eqref{induction} conserves}
\begin{equation}
    \Delta \psi = \int^{\infty}_{-\infty}\dd \tilde{\ell}\frac{\tilde{B}_0(\tilde{\ell})}{|\tilde{\bnabla} \alpha|^2} \tilde{J_{\parallel}} = \int^{\infty}_{-\infty}\dd \tilde{\ell} \tilde{B}_0(\tilde{\ell}) \delta \tilde{B}_{\perp}.\label{Deltapsi}
\end{equation}$\Delta \psi$ is the difference (``step'') in the coordinate $\psi$ of the perturbed field line between $\ell = -\infty$ and $\ell = +\infty$: $\Delta \psi \neq 0$ implies reconnection of field lines from inside the separatrix with those outside it [see Fig.~\ref{fig:illustration}(b)]. In a fusion device, this would cause the loss of confinement of hot plasma from inside the separatrix by streaming. \blue{The total flux reconnected through the separatrix [see Fig.~\ref{fig:illustration}(b)] is $\int \dd \alpha \dd \psi \sim \Delta \psi /n$.} Our goal is to determine the evolution of $\Delta \psi$ for finite $\mathcal{S}\gg 1$.

\textit{Late-time diffusive solutions.\quad}We consider the late-time ($\epsilon \tilde{t}\gg 1$) evolution of a diffusively spreading pulse induced by the force $\tilde{F}$. We plot numerical solutions of Eqs.~\eqref{vorticity} and~\eqref{induction} in Fig.~\ref{fig:solutions}. In what follows, we determine these solutions and their reconnection rates (Fig.~\ref{fig:Deltapsi}) \blue{for $\mathcal{S}\gg 1$  analytically, using matched asymptotic expansions. There are two relevant solution domains: an inner region, $\tilde{\ell}\ll \ln \mathcal{S}$, where inertia [left-hand side of Eq.~\eqref{vorticity}] is negligible, and an outer region, $\tilde{\ell}\gg \ln (\ln\mathcal{S})$, where induction [first term in Eq.~\eqref{induction}] is. We justify neglecting these terms in the above-stated domains \textit{a posteriori}---see Section~3 of the End Matter.}

\textit{MHD case:~(i) Inner region} \blue{($\tilde{\ell} \ll \ln \mathcal{S}$).\quad Neglecting inertia, Eq.~\eqref{vorticity} yields
\begin{equation}
    \tilde{J}_{\parallel\mathrm{in}}(\tilde{\ell},\tilde{t}) = - \int_{-\infty}^{\tilde{\ell}}  \dd \tilde{\ell}' \tilde{F} + \tilde{J}_{\parallel 0}(\tilde{t});\label{JsolninnerMHD}
\end{equation}substituting this into~\eqref{induction} yields, for $\epsilon \tilde{t}\gg 1$,
\begin{equation}
    \tilde{u}_{\mathrm{in}} = \frac{1}{\tilde{B}_0}\int_0^{\tilde{\ell}} \dd \tilde{\ell} ' \frac{\tilde{B}_0}{\mathcal{S}}\left( \frac{\mathcal{S}}{|\tilde{\bnabla} \alpha|^2}\frac{\dd \tilde{J}_{\parallel 0}}{\dd \tilde{t}} + \tilde{J}_{\parallel 0}-\int^{\ell'}_{-\infty}\dd \ell'' \tilde{F}\right).\label{usolninnerMHD}
\end{equation}}The inner solution depends on a single unknown function of time, $\tilde{J}_{\parallel 0}(t)$, which we shall determine by matching to the outer solution. We can deduce the value of $\tilde{J}_{\parallel 0}(\tilde{t})$ at early times by assuming that the forcing-ramp-up time $\epsilon^{-1}$ is small compared with the normalized reconnection timescale $\tau_{\mathrm{rec}}$, whatever the latter turns out to be (we show how to relax this restriction in Section~2 of the Supplementary Information). In that case, $\Delta \psi $ is unchanged between $\tilde{t}=0$ and $\epsilon^{-1} \ll \tilde{t}\ll \tau_{\mathrm{rec}}$. At such times,  Eqs.~\eqref{Deltapsi} and \eqref{JsolninnerMHD} require that
\begin{equation}
    \tilde{J}_{\parallel 0}(\tilde{t}) = \mathcal{J}_0 \equiv  |\Delta'|\int_{-\infty}^{\infty} \dd \tilde{\ell}\frac{\tilde{B}_0(\tilde{\ell})}{|\tilde{\bnabla} \alpha|^2}\int_{-\infty}^{\tilde{\ell}}  \dd \tilde{\ell}' \lim_{\tilde{t}\to \infty}\tilde{F},\label{J0}
\end{equation}where ${{1}/{|\Delta'|}\equiv \int_0^{\infty}\dd \tilde{\ell}{\tilde{B}_0(\tilde{\ell})}/{|\tilde{\bnabla} \alpha|^2}}$.

\textit{MHD case:~(ii) Outer region} ($\tilde{\ell} \gg \ln (\ln \mathcal{S})$).\quad Dropping $\tilde{F}$ in Eq.~\eqref{vorticity} (as ${\tilde{\ell}\gg 1}$) and neglecting induction in Eq.~\eqref{induction}, $\tilde{u}_{\mathrm{out}}$ satisfies the diffusion equation
\begin{equation}
    \frac{\p \tilde{u}_{\mathrm{out}}}{\p \tilde{t}} = \frac{\mathcal{S}}{|   \tilde{\bnabla} \alpha |^2} \frac{\p^2 \tilde{u}_{\mathrm{out}}}{\p \tilde{\ell}^2}. \label{MHD_outer_u}
\end{equation}

\blue{We now specialize to the equilibrium~\eqref{twocurrent}. For ${\tilde{\ell}\gg 1}$, Eq.~\eqref{gradalpha} reads ${|\tilde{\bnabla} \alpha|^2 = \Lambda e^{2\tilde{\ell}}}$}, where ${\Lambda = \exp[2\lim_{\tilde{\ell}\to\infty}(\tilde{\ell}(\tilde{z})-\tilde{z})]/4}$ is obtained from Eq.~\eqref{dldz}. Eq.~\eqref{MHD_outer_u} then admits a similarity variable $\xi = {\Lambda}{e^{2\tilde{\ell}}}/{4\mathcal{S}\tilde{t}}$. Changing variables from $\tilde{\ell}$ to $\xi$, Eq.~\eqref{MHD_outer_u} becomes
\begin{equation}
    \tilde{t}\frac{\p \tilde{u}_{\mathrm{out}}(\xi,\tilde{t})}{\p \tilde{t}} = \xi\frac{\p^2 \tilde{u}_{\mathrm{out}}(\xi,\tilde{t})}{\p \xi^2} + (1+\xi)\frac{\p \tilde{u}_{\mathrm{out}}(\xi,\tilde{t})}{\p \xi},\label{MHD_outer_v}
\end{equation}of which the series
\begin{equation}
    \tilde{u}_{\mathrm{out}}(\xi,\tilde{t}) = \sum_n \tilde{t}^{p_n} f_n(\xi)\label{series_soln}
\end{equation}is a solution if $f_n(\xi)$ satisfies
\begin{equation}
    \xi\frac{\dd^2 f_n}{\dd \xi^2} + (1+\xi)\frac{\dd f_n}{\dd \xi}-p_n f_n=0, \quad \forall n.\label{TricomiEquationMHD}
\end{equation}The solutions of Eq.~\eqref{TricomiEquationMHD} that vanish at $\xi\to\infty$ are ${f_n(\xi) = C_n e^{-\xi} U\left(1 + p_n, 1, \xi\right),\label{TricomiSolnMHD}}$ where $C_n$ is a constant and $U$ is Tricomi's confluent hypergeometric function. 

\blue{\textit{MHD case:~(iii) Matching solutions.\quad}Both the inner and outer solutions are valid for ${1 \ll \ln(\ln \mathcal{S}) \ll \tilde{\ell} \ll \ln \mathcal{S}}$ and must therefore match there. In this range, $\tilde{u}$ decreases linearly with $\tilde{\ell}$, i.e., ${\tilde{u} \to  c_1(\tilde{t}) - c_2(\tilde{t}) \tilde{\ell}}$ (Fig.~\ref{fig:solutions}). Apart from at very late times $\tilde{t}\gtrsim \mathcal{S}$ [see ``\textit{MHD case~(v)}''], only the first term in Eq.~\eqref{usolninnerMHD} for $\tilde{u}_{\mathrm{in}}$ contributes to $c_1$; the other two (resistive) terms determine $c_2$. From the outer solution, $c_1 = \lim_{\tilde{\ell}\to 0} \tilde{u}_{\mathrm{out}}$. Equating the two expressions for $c_1$, the matching condition is
\begin{equation}
    \lim_{\tilde{\ell} \to 0} \tilde{u}_{\mathrm{out}} =  \frac{\mathcal{S}}{|\Delta '|}\frac{\dd}{\dd\tilde{t}}\lim_{\tilde{\ell}\to 0}\frac{\p \tilde{u}_{\mathrm{out}}}{\p \tilde{\ell}},\label{matchingcond}
\end{equation}where we have used that $\tilde{J}_{\parallel 0}=\mathcal{S}\p \tilde{u}_{\mathrm{out}}/\p \tilde{\ell}$ in the matching region. Using the above-stated solutions of Eq.~\eqref{TricomiEquationMHD} and standard results for limits of Tricomi's function, Eq.~\eqref{series_soln} implies that}
\begin{equation}
    \lim_{\xi \to 0}\tilde{u}_{\mathrm{out}}
 = - \sum_n \frac{C_n\tilde{t}^{p_n}\ln (\Theta_n \xi)}{\Gamma(1+p_n)} + O(\xi),\label{MHD_limx0_u}
\end{equation}where $\Theta_n \equiv e^{2 \gamma_\mathrm{E} + \mathrm{F}\left(1 + n\right)}$, $\gamma_\mathrm{E}$ is the Euler--Mascheroni constant and $\mathrm{F}$ is the digamma function. For $\tilde{t}\ll \mathcal{S}$, the logarithm in Eq.~\eqref{MHD_limx0_u} is constant and approximately equal to $\ln (\Lambda/4\mathcal{S})$. Eqs.~\eqref{matchingcond} and~\eqref{MHD_limx0_u} together require that $p_{n+1}=p_n+1$, $C_{n+1}=-C_{n}/\tau_{\mathrm{rec}}$, where
\begin{equation}
    \tau_{\mathrm{rec}}= \frac{2\mathcal{S}}{|\Delta'| \ln (4\mathcal{S}/\Lambda)}.\label{MHDrecrate}
\end{equation}Finally, Eq.~\eqref{J0} requires $p_0=0$, $C_0 = -\mathcal{J}_0/2\mathcal{S}$, whence
\begin{align}
    \tilde{u}_{\mathrm{out}}&=-\frac{\mathcal{J}_0}{2\mathcal{S}}\sum^{\infty}_{n=0} (-1)^n e^{-\xi}U(1+n,1,\xi) \left(\frac{\tilde{t}}{\tau_{\mathrm{rec}}}\right)^n,    \label{usolnouterMHD}\\\tilde{J}_{\parallel\mathrm{out}}&=\mathcal{J}_0\sum^{\infty}_{n=0} (-1)^n  \xi e^{-\xi}U(1+n,2,\xi) \left(\frac{\tilde{t}}{\tau_{\mathrm{rec}}}\right)^n.\label{JsolnouterMHD}
\end{align}The inner solution is given by Eqs.~\eqref{JsolninnerMHD} and~\eqref{usolninnerMHD}, with $\tilde{J}_{\parallel 0}$ the $\xi\to 0$ limit of Eq.~\eqref{JsolnouterMHD}, i.e.,
\begin{equation}
    \tilde{J}_{\parallel 0}=\sum^{\infty}_{n=0} \frac{(-1)^n}{n!} \left(\frac{\tilde{t}}{\tau_{\mathrm{rec}}}\right)^n = \mathcal{J}_0\exp \left(-\frac{\tilde{t}}{\tau_{\mathrm{rec}}}\right).\label{Jp0solnMHD}
\end{equation}Eqs.~\eqref{JsolninnerMHD},~\eqref{usolninnerMHD},~\eqref{usolnouterMHD}, and~\eqref{JsolnouterMHD} are as we obtain from numerical solution of Eqs.~\eqref{vorticity} and~\eqref{induction} (see Fig.~\ref{fig:solutions}). Qualitatively, the solution for $\tilde{t}\ll \tau_{\mathrm{rec}}$ [the $n=0$ terms in Eqs.~\eqref{usolnouterMHD} and~\eqref{JsolnouterMHD}] is a self-similar spreading with similarity variable $\xi$. The solution is diffusive, but corresponds to negligible reconnection [Eq.~\eqref{Deltapsi_MHD}] until the current starts to decay at $\tilde{t}\sim \tau_{\mathrm{rec}}$, which is when the propagating front reaches $\tilde{\ell}\sim \ln \mathcal{S}$ (corresponding to~$\xi \sim 1$).

\textit{MHD case:~(iv) Reconnection rate.\quad}It follows from Eqs.~\eqref{Deltapsi},~\eqref{JsolninnerMHD} and~\eqref{Jp0solnMHD} that
\begin{equation}
    \Delta \psi = \frac{1}{|\Delta'|} (\tilde{J}_{\parallel 0} - \mathcal{J}_{0}) = -\frac{\mathcal{J}_{0}}{|\Delta'|} \left[1 - \exp \left(-\frac{\tilde{t}}{\tau_{\mathrm{rec}}}\right)\right]\label{Deltapsi_MHD}
\end{equation}which confirms that $\tau_{\mathrm{rec}}$ [Eq.~\eqref{MHDrecrate}] is the reconnection timescale. Eq.~\eqref{Deltapsi} correctly reproduces the evolution of $\Delta \psi$ that we obtain numerically (see Fig.~\ref{fig:Deltapsi}).

\textit{MHD case:~(v) Matching solutions, $\tilde{t}\gtrsim \mathcal{S}$.\quad}Combining Eqs.~\eqref{usolninnerMHD} and~\eqref{Jp0solnMHD}, we find that the contribution of the final term (involving $\tilde{F}$) in~\eqref{usolninnerMHD} to  $c_1(\tilde{t})$ is negligible only for $\tilde{t}\ll \mathcal{S}$---at later times, this term must also be included in the matching condition~\eqref{matchingcond}. Physically, at $\tilde{t}\sim \mathcal{S}$, resistivity can no longer be neglected at the location of the force, and the flux tube begins to ``slip'' diffusively there. We derive the corresponding solution in Section 4 of the End Matter~\footnote{Qualitatively, the effect of the modification is that decay of $\tilde{u}_{\mathrm{out}}$ and $\tilde{J}_{\parallel\mathrm{out}}$ is arrested at a non-zero value.}. However, because $\tau_{\mathrm{rec}}\ll \mathcal{S}$ as $\mathcal{S}\to\infty$, the solution in the preceding sections is the one of primary interest.

\blue{\textit{Kinetic case:\quad}The derivation of the reconnection rate in the kinetic case of Eq.~\eqref{vorticity} is analogous to the MHD case; we present it in Section~2 of the End Matter. The chief difference is that, unlike Eq.~\eqref{MHD_outer_u}, the diffusion equation for the outer region in the kinetic case [Eq.~\eqref{kinetic_diffusion}] is independent of $|\bnabla \alpha|$, so the diffusion coefficient is not suppressed by field-line separation. We find~[cf.~Eq.~\eqref{Deltapsi_MHD}]}
\begin{equation}
    \Delta \psi = -\frac{\mathcal{J}_{0}}{|\Delta'|} \left[1 - E_{3/2}\left(-\left(\frac{\tilde{t}}{\tau_{\mathrm{rec}}}\right)^{3/2}\right)\right],\label{Deltapsi_kinetic}
\end{equation}where $\tau_{\mathrm{rec}}=(\mathcal{S}/|\Delta'|^2)^{1/3}$ and $E_{3/2}(x)$ is the Mittag--Leffler function, a generalization of the exponential function defined by ${E_{\alpha}(x)=\sum_{k=0}^{\infty} x^k/\Gamma(1+\alpha k)}$.

\textit{Generalization to other geometries.\quad}As explained above, the outer-region diffusion equation is independent of~$|\tilde{\bnabla} \alpha|(\ell)$ in the kinetic case. The solution~\eqref{Deltapsi_kinetic} is therefore valid for \emph{any} magnetic geometry, not just the one described by Eq.~\eqref{twocurrent}. By contrast, $|\tilde{\bnabla} \alpha|$ does appear in Eq.~\eqref{MHD_outer_v} (MHD case). For an equilibrium with $|\tilde{\bnabla} \alpha|\propto \tilde{\ell}^{\delta}$ for $\tilde{\ell}\gg 1$, it is readily verified that Eq.~\eqref{MHD_outer_v} has a similarity variable $\xi=\tilde{\ell}/(\mathcal{S}\tilde{t})^{1/2(\delta +1)}$. It follows that~\footnote{In deriving these expressions, we have used the relations ${\tilde{J}_{\parallel 0} = \mathcal{S} \lim_{\tilde{\ell} \to 0} \p \tilde{u}_{\mathrm{out}}/\p \tilde{\ell}}$, ${\dd \tilde{J}_{\parallel 0}/\dd \tilde{t} = \lim_{\tilde{\ell} \to 0} \tilde{u}_{\mathrm{out}}}$ and ${\lim_{\tilde{t}\to 0} \tilde{u}_{\mathrm{out}}\propto \tilde{t}^{p_0}f_0(\xi)}$ [Eq.~\eqref{series_soln}], which together imply that ${\dd \ln J_{\parallel 0}/\dd \tilde{t} \propto \blue{\mathcal{S}^{-(1+2\delta)/2(1+\delta)}} \tilde{t}^{1/2(\delta+1)}}$ as $\tilde{t}\to 0$}
\begin{equation}
    \Delta \psi \to  \left( \frac{\tilde{t}}{\tau_{\mathrm{rec}}}\right)^{\tfrac{\delta+3/2}{\delta+1}}, \quad\tau_{\mathrm{rec}}\propto \frac{\mathcal{S}^{(2\delta + 1)/(2\delta+3)}}{|\Delta '|^{2(\delta+1)/(2\delta+3)}}\label{generalscaling}
\end{equation}as $\tilde{t}/\tau_{\mathrm{rec}}\to 0$. Eq.~\eqref{generalscaling} implies that MHD reconnection becomes \textit{slower} as the rate of field-line separation increases (i.e., as $\delta$ increases). In the case of linear shear, i.e., $\delta = 1$ \blue{(as for field lines \emph{not} on the separatrix)} $\tau_{\mathrm{rec}}\propto \mathcal{S}^{3/5}$~\footnote{This dependence was inevitable as, in the case of linear shear, Eqs.~\eqref{vorticity} and~\eqref{induction} are isomorphic to the 2D tearing-mode equations~\cite{Roberts65}.}. \blue{Faster reconnection occurs for field lines that separate more slowly (decreasing $\delta$), but only up to $\delta = 1/2$, as then $|\Delta'|$ diverges [see its definition below~\eqref{J0}]. Physically, $\delta=1/2$ is the smallest value of $\delta$ for which the diffusive solution can be causally connected to the origin at $t\sim \tau_{\mathrm{rec}}$. The shortest possible MHD-reconnection timescale in our theory as $\mathcal{S}\to \infty$ is therefore $\tau_{\mathrm{rec}}\sim\mathcal{S}^{1/2}/|\Delta|^{3/4}$, corresponding to~$\delta \to 1/2^{+}$.}

\textit{Conclusion.\quad}
We have derived nonlinear solutions for 3D MHD and electron-only reconnection in a sheared magnetic field. In our solutions, reconnection is induced by diffusing Alfvén- and semi-collisional kinetic-Alfvén-wave packets generated by an external force applied gradually to exponentially separating field lines. Their diffusion changes the connectivity of the global field in timescales $\propto \mathcal{S}/\ln \mathcal{S}$ and $\propto\mathcal{S}^{1/3}$, respectively. \blue{We expect our results to be quantitatively applicable to reconnection at separatrices, as in tokamak-edge-field schochasticization by an ELM-control (RMP) coil, although we note that the anisotropic ordering $k_{\parallel}\gg k_{\perp}$ necessary for a tractable theory is not rigorously applicable in that case. Qualitatively, our paper provides a solvable case of 3D magnetic reconnection, with the interesting and unexpected property that the reconnection rate is greater for field lines that diverge more slowly [Eq.~\eqref{generalscaling}] for MHD dynamics, and is geometry-independent in the case of semi-collisional electron-only reconnection~\cite{ZoccoSchekochihin11}.}

\vspace{.6cm}

\textit{Acknowledgements.\quad}We are grateful to Christopher Ham and Anna Thackray for their collaboration on a related project. We also thank the hosts, organizers, and participants of the 2024 Plasma Kinetics Workshop at the Wolfgang Pauli Institute, Vienna, where this work was discussed.

\newpage
\clearpage

\newpage

\setcounter{page}{1}

\section*{End matter}

\subsection{1. Derivation of $|\mathbf{\nabla} \alpha|^2 $ and $B_0(\ell)$ for the field~\eqref{twocurrent}}

From the vector product of $\bB_0=\bnabla 
\psi\times\bnabla\alpha$ with $\bnabla \psi$,
\begin{equation}
\bnabla\alpha = \frac{\bB_0\times\nabla\psi}{|\nabla\psi|^2} + \lambda\nabla\psi,\;\; \lambda = -\int_{L}^\ell\frac{d\ell'}{B_0}{\cal I}(\psi, \alpha,\ell'),
\label{localS}
\end{equation}where \blue{the lower limit} $L = L(\alpha, \psi)$ is arbitrary and 
\begin{equation}
{\cal I}(\psi, \alpha,\ell) = \frac{\bB_0\times\bnabla\psi}{|\bnabla\psi|^2}\bcdot\bnabla\times\frac{{\bB_0}\times\bnabla\psi}{|\bnabla\psi|^2}
\label{localS2}\end{equation} 
is the local shear. It follows from~\eqref{localS} that
\begin{equation}
    |\bnabla \alpha|^2 = \frac{B_0^2}{|\bnabla \psi|^2}+\lambda^2 |\bnabla \psi|^2.\label{gradalpha2}
\end{equation}We proceed to evaluate~\eqref{gradalpha2} on the separatrix of the magnetic field~\eqref{twocurrent}. Substituting~\eqref{twocurrent} into~\eqref{localS2}, we obtain
\begin{equation}
{\cal I} = -\frac{B_{\mathrm{ext}}\bnabla\psi\bcdot\bnabla|\bnabla\psi|^2}{a|\bnabla\psi|^4}  
 = -\frac{\ell_0}{a^4}\left(3 + \frac{(a^4 - \chi)}{r^4}\right)
\label{localS3}
\end{equation}where ${\chi \equiv \exp(c\psi /a I_0) = (r^2+a^2)^2 -4a^2 x^2}$ and we have used the fact that $|\bnabla\chi|^2 = 16r^2\chi$. On the separatrix, where $\chi = a^4$, ${\cal I} = -3\ell_0/a^4$, whence
\begin{equation}
 \frac{|\bnabla\psi|^2}{B_{\mathrm{ext}}^2} = \frac{a^2r^2}{\ell_0^2}, \quad  \frac{B_0^2}{B_{\mathrm{ext}}^2} = 1 + \frac{r^2}{\ell_0^2}, \quad \lambda = \frac{3\ell_0 z}{B_{\mathrm{ext}} a^4}.
\label{gradpsi}
\end{equation}

It remains to determine $r(\ell)$. $\chi$ is constant along field lines, so the $x$ and $y$ components of field-line trajectories are ${x(r, \chi) = \pm (1/(2a))[\sqrt{((r^2+a^2)^2 - \chi)}]}$ and ${y(r, \chi) = \pm (1/(2a))[\sqrt{(\chi - (r^2-a^2)^2)}]}$, respectively. We determine the field-line trajectory in $z$ by integrating $(\p z/ \p r)_{\psi, \alpha}= (\bB_0\bcdot \bnabla z) / (\bB_0\bcdot \bnabla r)$ to yield (in the quadrant $y>0, x<0$)
\begin{equation}
\frac{z-a\alpha}{2l_0a^2} = \frac{\chi}{a^4}\int^r_{r_0}\frac{rdr}{\sqrt{((r^2+a^2)^2 - \chi)(\chi - (r^2-a^2)^2)}}.
\label{fieldlinetwo1}
\end{equation}We have fixed the constant of integration in~\eqref{fieldlinetwo1} by choosing $\alpha = z/a$ on the semi-infinite surface ${y=0}$, ${x<-a}$ [see Fig.~\ref{fig:illustration}(b)]. In general, Eq.~(\ref{twocurrent}) requires $a\alpha = z + g(\chi, r) + f(\chi)$, where $g(\chi,r)$ is the function that generates the constant field~$B_{\mathrm{ext}} \hat{\bz}$ and $f(\chi)$ is arbitrary---our choice corresponds to $f(\chi) = -g(\chi, r_0(\chi))$ with $r_0(\chi) = {\sqrt{a^2 + \sqrt{\chi}}}$. Evaluating the integral~\eqref{fieldlinetwo1} with $\chi = a^4$ and (without loss of generality) $\alpha = 0$, we find
\begin{equation}
r^2 = \frac{2a^2}{\cosh (2z/\ell_0)}.
\label{sepfieldline}
\end{equation}Eqs.~\eqref{gradalpha} and~\eqref{dldz} in the main text follow from substitution of~\eqref{sepfieldline} into~\eqref{gradpsi} and~\eqref{gradalpha2}.

\color{black}

\vspace{-3mm}
\subsection{2. Solution for the kinetic case}
\vspace{-2mm}

\textit{Kinetic case:~(i) Inner region $(\tilde{\ell}\ll \mathcal{S}^{2/3})$.\quad}We neglect inertia in the inner region (as justified \textit{a posteriori} in Section 3 of the End Matter). This is the only term in which the kinetic and MHD cases of Eq.~\eqref{vorticity} differ, so Eqs.~\eqref{JsolninnerMHD} and~\eqref{usolninnerMHD} remain valid for the kinetic case.

\textit{Kinetic case:~(ii) Outer region $(\tilde{\ell}\gg \ln \mathcal{S})$.\quad}
As justified \textit{a posteriori} in Section 3 of the End Matter, we now neglect the forcing term in Eq.~\eqref{vorticity} and the induction term in Eq.~\eqref{induction}. Then, $\tilde{u}_{\mathrm{out}}$ obeys the diffusion equation [cf. Eq.~\eqref{MHD_outer_u} for the MHD case]
\begin{equation}
    \frac{\p \tilde{u}_{\mathrm{out}}}{\p \tilde{t}} = {\mathcal{S}} \frac{\p^2 \tilde{u}_{\mathrm{out}}}{\p \tilde{\ell}^2}. \label{kinetic_diffusion}
\end{equation}We recast Eq.~\eqref{kinetic_diffusion} in a form similar to Eq.~\eqref{MHD_outer_v} by changing variables from $\tilde{\ell}$ to a new similarity variable ${\xi = \tilde{\ell}/(4\mathcal{S}\tilde{t})^{1/2}}$, which yields
\begin{equation}
    \tilde{t}\frac{\p \tilde{u}_{\mathrm{out}}}{\p \tilde{t}} = \frac{1}{4}\frac{\p^2 \tilde{u}_{\mathrm{out}}}{\p \xi^2} + \frac{1}{2}\xi \frac{\p \tilde{u}_{\mathrm{out}}}{\p \xi}\label{kinetic_outer_u}.
\end{equation}The solutions of the ordinary differential equations that result from substituting Eq.~\eqref{series_soln} into Eq.~\eqref{kinetic_outer_u} 
are ${f_n(\xi) =  C_n e^{-\xi^2} U\left((1+2p_n)/2,1/2,\xi^2\right)}$.

\begin{figure}
    \centering
    \includegraphics[width=\columnwidth]{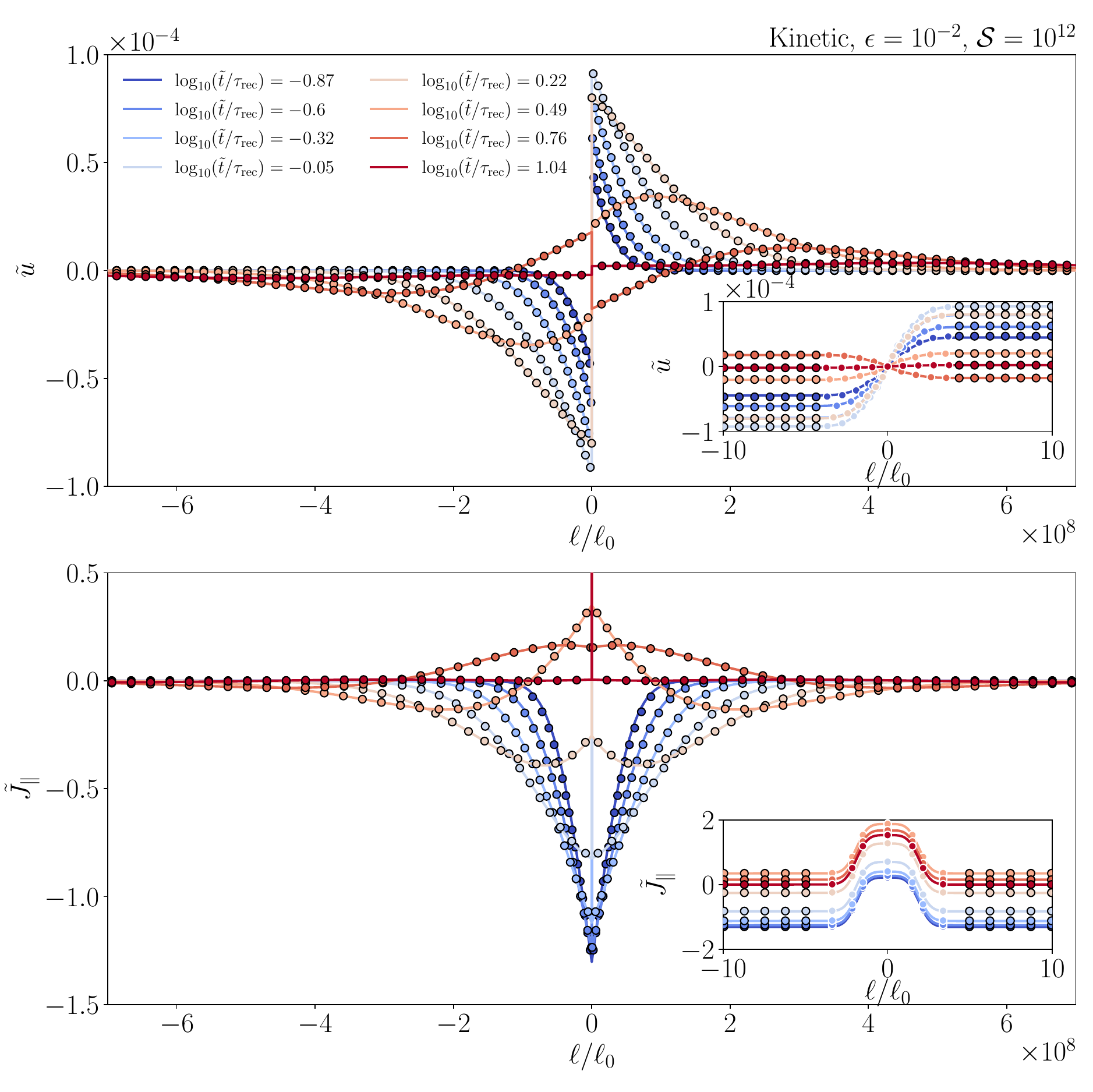}
    \caption{\blue{As Fig.~\ref{fig:solutions}, but for the kinetic case.}}
    \label{fig:kinetic_solutions}
\end{figure}

\textit{Kinetic case:~(iii) Matching solutions.\quad}Because Eq.~\eqref{induction} is the same in the MHD and kinetic cases, Eq.~\eqref{matchingcond} is the matching condition for both [see below Eq.~\eqref{Jp0_kinetic} for justification of neglect of the second two terms in Eq.~\eqref{usolninnerMHD} in forming the matching condition]. From the small-argument limit of Tricomi's function, we have
\begin{equation}
    \lim_{\xi\to 0}\tilde{u}_{\mathrm{out}} = \sqrt{\pi} \sum_n C_n \tilde{t}^{p_n}\left[
\frac{1}{\Gamma\left(1 + p_n\right)} 
- \frac{2\xi}{\Gamma\left(1/2+p_n\right)} \right]\label{kinetic_limx0_u}
\end{equation}up to $O\left(\xi^{2}\right)$. Eq.~\eqref{matchingcond} then requires that $p_{n+1}=p_n + 3/2$ and gives the recursion relation $C_{n+1}=-|\Delta'|/\mathcal{S}^{1/2}C_n$, so $C_{n}=(-|\Delta'|/\mathcal{S}^{1/2})^n C_0$. Matching to the early-time solution~\eqref{J0} gives $p_0 = 1/2$, $C_0 = -\mathcal{J}/\sqrt{\pi}\mathcal{S}^{1/2}$, whence,
\begin{align}
    \tilde{u}_{\mathrm{out}}&=-\mathcal{J}_0 \sqrt{\frac{ \tilde{t}}{\pi \mathcal{S}} }\sum^{\infty}_{n=0} (-1)^n e^{-\xi^2}  U_{n,1/2}(\xi)\left(\frac{\tilde{t}}{\tau_{\mathrm{rec}}}\right)^{3n/2},
    \label{usolnouterkinetic}\\\tilde{J}_{\parallel\mathrm{out}}&=\frac{\mathcal{J}_0}{\sqrt{\pi}}\sum^{\infty}_{n=0} (-1)^n e^{-\xi^2}  \xi U_{n,3/2}(\xi)\left(\frac{\tilde{t}}{\tau_{\mathrm{rec}}}\right)^{3n/2},\label{Jsolnouterkinetic}
\end{align}where 
\begin{equation}
    \tau_{\mathrm{rec}}=\left(\frac{\mathcal{S}}{|\Delta'|^2}\right)^{1/3},\quad {U_{n,\alpha}\equiv U(1+3n/2,\alpha,\xi^2)}.\label{tau_kinetic}
\end{equation}These solutions are as we find numerically (Fig.~\ref{fig:kinetic_solutions}). Qualitatively, both $\tilde{u}_{\mathrm{out}}$ and $\tilde{J}_{\parallel\mathrm{out}}$ spread self-similarly for ${\tilde{t}\ll \tau_{\mathrm{rec}}}$, with $\tilde{u}_{\mathrm{out}}$ also growing like~$\tilde{t}^{1/2}$. Reconnection occurs when the current begins to decay at $\tilde{t}\sim \tau_{\mathrm{rec}}$, which is when the propagating front reaches~${\tilde{\ell}\sim \mathcal{S}^{2/3}}$. The ${\xi\to 0}$ limit of Eq.~\eqref{Jsolnouterkinetic} yields
\begin{equation}
    \tilde{J}_{\parallel 0} =\mathcal{J}_{0} E_{3/2}\left[-\left(\tilde{t}/\tau_{\mathrm{rec}}\right)^{3/2}\right],\label{Jp0_kinetic}
\end{equation}where $E_{3/2}(x)$ is the Mittag--Leffler function. We note that $\tilde{J}_{\parallel 0} = O[ (\tilde{t}/\tau_{\mathrm{rec}})^{-3/2}]$ as $\tilde{t}/\tau_{\mathrm{rec}}\to \infty$, from which it follows that the term involving $\tilde{F}$ in Eq.~\eqref{usolninnerMHD} does not contribute to $c_1$ [see ``\textit{MHD case:~(iii)}''] until ${\tilde{t} \sim \mathcal{S} \gg \tau_{\mathrm{rec}}}$, and therefore is always negligible in forming the matching condition~\eqref{matchingcond} [cf. ``\textit{MHD case:~(iv)}''].  Combining Eqs.~\eqref{Deltapsi} and~\eqref{Jp0_kinetic} yields Eq.~\eqref{Deltapsi_kinetic} for $\Delta \psi$.

\color{black}

\vspace{-3mm}
\subsection{3. \textit{A posteriori} justification of neglected terms in matched asymptotics}
\vspace{-2mm}

\textit{MHD case: Inner region.} From Eqs.~\eqref{JsolninnerMHD} and~\eqref{usolninnerMHD}, we have that, for $\epsilon \tilde{t}\gg 1$ and $\tilde{\ell}\gg 1$,
\begin{equation}
    \frac{\p \tilde{u}_{\mathrm{in}}}{\p \tilde{t}}= \frac{1}{B_0}\int_{0}^{\ell}\dd \ell'\frac{B_0}{|\bnabla \alpha|^2} \left(\frac{\dd ^2 \tilde{J}_{\parallel 0}}{\dd \tilde{t}^2}+ \frac{|\bnabla \alpha|^2}{\mathcal{S}}\frac{\dd \tilde{J}_{\parallel 0}}{\dd \tilde{t}}{}\right).\label{dudt}
\end{equation}
Using Eq.~\eqref{Jp0solnMHD} for $\tilde{J}_{\parallel 0}$ in Eq.~\eqref{dudt}, the neglected term involving $\p \tilde{u}_{\mathrm{in}}/\p \tilde{t}$ in Eq.~\eqref{JsolninnerMHD} is small compared to $\tilde{J}_{\parallel 0}$ if
\begin{equation}
    \int_{-\infty}^{\tilde{\ell}}\frac{|\tilde{\bnabla} \alpha|^2}{B_0} \left(\frac{1}{|\Delta'|\tau_{\mathrm{rec}}} - \frac{\tilde{\ell}}{\mathcal{S}}
    \right) \frac{1}{\tau_{\mathrm{rec}}} \ll 1.\label{inner_condition}
\end{equation}This is true for $\tilde{\ell}\ll \ln \mathcal{S}$, which is therefore the domain of validity of the inner solution.

\textit{MHD case: Outer region.} The induction term in Eq.~\eqref{induction} is negligible compared to the diffusive one for sufficiently small~$\tilde{\ell}$ because of its $1/|\tilde{\bnabla} \alpha|^2$ prefactor [from Eq.~\eqref{JsolnouterMHD} (visualized in Fig.~\ref{fig:solutions}), $\p \ln \tilde{J}_{\parallel \mathrm{out}}/\p t$ has no strong dependence on $\tilde{\ell}$ to prevent this]. Assuming that the outer solution is valid for $\tilde{\ell}\ll \ln \mathcal{S}$, then matching to the inner condition yields $\tilde{J}_{\parallel \mathrm{out}} = \tilde{J}_{\parallel 0}$ there. It follows that the ratio of the induction to resistive terms, $({\mathcal{S}/|\tilde{\bnabla} \alpha|^2})\p \ln \tilde{J}_{\parallel \mathrm{out}}/\p t $, is $\ln \mathcal{S}/|\tilde{\bnabla} \alpha|^2$ at $\tilde{\ell} \ll \ln \mathcal{S}$. The induction term is therefore negligible for $\tilde{\ell}\gg \ln (\ln \mathcal{S})$, which is the domain of validity of the outer solution (this includes $\tilde{\ell} \ll \ln \mathcal{S}$, as assumed).

\textit{Kinetic case: Inner region.}  Eq.~\eqref{dudt} holds for the kinetic case; substituting Eq.~\eqref{Jp0_kinetic} and using Eq.~\eqref{tau_kinetic} for~$\tau_{\mathrm{rec}}$, it follows from an analogous condition to~\eqref{inner_condition} that $\p \tilde{u}_{\mathrm{in}}/\p \tilde{t}$ is negligible in Eq.~\eqref{JsolninnerMHD} for ${\tilde{\ell}\ll \mathcal{S}^{2/3}}$. This is the domain of validity of the inner solution. 

\textit{Kinetic case: Outer region.} The ratio of induction and diffusion terms in~\eqref{induction} is $\mathcal{S}/(\tau_{\mathrm{rec}}|\bnabla \alpha|^2)$. This is small provided that $\ell \gg \ln \mathrm{S}$, which is therefore the domain of validity of the outer solution.

\color{black}

\vspace{-3mm}
\subsection{4. MHD solution for $\tilde{t}\sim \mathcal{S}$}
\vspace{-2mm}

Retaining all terms in Eq.~\eqref{usolninnerMHD}, Eq.~\eqref{matchingcond} becomes, for~$\epsilon \tilde{t}\gg 1$ [using Eq.~\eqref{JsolninnerMHD} and ${\tilde{J}_{\parallel \mathrm{out}}=\mathcal{S}\p \tilde{u}_{\mathrm{out}}/\p \tilde{\ell}}$],
\begin{equation}
    \lim_{\tilde{\ell} \to 0} \tilde{u}_{\mathrm{out}}  =  \frac{\mathcal{S}}{|\Delta '|}\frac{\dd}{\dd\tilde{t}}\lim_{\tilde{\ell}\to 0}\frac{\p \tilde{u}_{\mathrm{out}}}{\p \tilde{\ell}} + \tilde{u}_{F\mathcal{S}} + \zeta(\tilde{\ell}) \lim_{\tilde{\ell} \to 0 }\frac{\p \tilde{u}_{\mathrm{out}}}{\p \tilde{\ell}},\label{MHDtgtrSmatching}
\end{equation}where $\zeta(\tilde{\ell})\equiv\tilde{B}_0(\tilde{\ell})^{-1}\int_0^{\tilde{\ell}} \dd \tilde{\ell}' \tilde{B}_0(\tilde{\ell} ')$ and
\begin{equation}
    \tilde{u}_{F\mathcal{S}} \equiv -\frac{1}{\mathcal{S}} \int^\infty_0 \dd \tilde{\ell}' \tilde{B}_0(\tilde{\ell}')\int^{\tilde{\ell} '}_{-\infty} \dd \tilde{\ell} '' F(\tilde{\ell}'').
\end{equation}Substituting Eq.~\eqref{MHD_limx0_u} into Eq.~\eqref{MHDtgtrSmatching} yields
\begin{multline}
    \sum_n  \frac{C_n \tilde{t}^{p_n}(\ln (4\mathcal{S}e^{2\zeta_0}/\Theta_n\Lambda)+\ln \tilde{t})}{\Gamma(1+p_n)} \\= \tilde{u}_{F\mathcal{S}} -\frac{\mathcal{S}}{|\Delta'|} \sum_n \frac{2C_np_n\tilde{t}^{p_n-1}}{\Gamma(1+p_n)},
\end{multline}where $\zeta_0 = \lim_{\tilde{\ell} \to \infty} [\int_0^{\tilde{\ell}} \dd \tilde{\ell}' \tilde{B}_0(\tilde{\ell}') - \tilde{\ell}]$. Expanding around $\tilde{t}=\tilde{t}_0$ for some arbitrary $\tilde{t}_0>0$, we obtain
\begin{align}
    \sum_n  \bar{C}_n \tau_0(1+\delta \bar{\tau})^{n+1}[\ln (\bar{\mathcal{S}}\tau_0/\Theta_n)+\ln (1+\delta \bar{\tau})] \nonumber\\= \tilde{u}_{F\mathcal{S}} \tau_0 (1 + \delta\bar{\tau})-\sum_n {\bar{C}_n n(1+\delta \bar{\tau})^{n}},\label{matrixCn}
\end{align}where for convenience we have defined $\tau = \tilde{t}/(2\mathcal{S}/|\Delta|')$, ${\tau_0 = \tilde{t}_0/(2\mathcal{S}/|\Delta|')}$, ${\delta\bar{\tau}= (\tau-\tau_0) / \tau_0}$, ${\bar{\mathcal{S}}=8 \mathcal{S}^2 e^{2\zeta_0}/ \Lambda|\Delta'|}$ and ${\bar{C}_n = C_n ({2\mathcal{S}\tau_0/|\Delta'|})^n/\Gamma[(n+2)/2]}$. To obtain~$\bar{C}_n$, we truncate the sums (at 35 terms) and expand~\eqref{matrixCn} in powers of $\delta \bar{\tau}$. Equating coefficients of each power of $\delta \bar{\tau}$ yields a dense matrix equation for~$\bar{C}_n$, which we solve numerically. The dots in Fig.~\ref{fig:solutions} (left panels) show the corresponding solutions. In constructing these solutions, we choose $\tilde{t}_0$ to be equal to the time of evaluation.

\color{black}

\newpage
\clearpage

\newpage

\setcounter{page}{1}

\section*{Supplementary information}

\vspace{-3mm}
\subsection{1. Derivation of the dynamical equations}
\vspace{-2mm}

We derive Eqs.~\eqref{vorticity} and~\eqref{induction} in the main text from
\begin{equation}
    \frac{\p }{\p t }\hat{G}\Phi + \{\Phi,\hat{G}\Phi\} = v_{A}(\ell) \frac{\p}{\p \ell}\nabla_\perp^2 \Psi + \{\Psi,\nabla_\perp^2\Psi\}\label{RMHD1}
\end{equation}and
\begin{equation}
    \frac{\p }{\p t }\Psi + \{\Phi,\Psi\} = \frac{\p}{\p \ell}\left[v_{A}(\ell) \Phi\right] + \eta \nabla_\perp^2\Psi,\label{RMHD2}
\end{equation}where $\Phi$ is the streamfunction for the velocity field perpendicular to the local unperturbed magnetic field, ${\bu_{\perp}=\bb_0\times\bnabla_{\perp}\Phi}$; $\Psi$ is the flux function for the magnetic perturbation, ${\delta\bB_{\perp}=\sqrt{4\pi \rho_0}\bb_0\times\bnabla_{\perp}\Psi}$; ${v_A(\ell)=B_0(\ell)/\sqrt{4\pi\rho_0}}$ is the local Alfv\'{e}n speed associated with the unperturbed field strength $B_0(\ell)$ and density $\rho_0$; the Poisson bracket is
\begin{equation}
    \{\Phi, \Psi\} \equiv \bb_0 \bcdot (\bnabla_\perp \Phi \times \bnabla_\perp \Psi);\label{PoissonBracket1}
\end{equation}$\bb_0$ is the unit vector in the direction of $\bB_0$ [Eq.~\eqref{Beqn}];
\begin{equation}
    \hat{G}=\frac{2}{\rho_i^2}(\hat{\Gamma}_0-1),\label{Goperator}
\end{equation}where $\rho_i$ is the ion gyroradius; $\hat{\Gamma}_0$ is the operator whose representation in Fourier space is as multiplication by
\begin{equation}
    \Gamma_0(\alpha_i) \equiv I_0(\alpha_i) e^{-\alpha_i},\quad \alpha_i \equiv \frac{k_{\perp}^2\rho_i^2}{2};
\end{equation}and $I_0$ is the modified Bessel function of the first kind. 

Equations \eqref{RMHD1} and \eqref{RMHD2} are the semi-collisional limit of the Kinetic Reduced Electron Heating Model (KREHM) of Ref.~\cite{ZoccoSchekochihin11}. They describe the low-frequency (compared with the ion gyrofrequency) anisotropic ($ k_{\parallel}/k_{\perp}\ll1$, with $k_{\parallel}$ and $k_{\perp}$ the wavenumbers along and across the equilibrium magnetic field, respectively) dynamics of a semi-collisional plasma (collisional electrons but collisionless ions) at small plasma $\beta$ and large ion-to-electron temperature ratio. For that system, ions have no mean flow and the magnetic field is advected by the $\bE\times\bB$ motion of electrons. In this study, we focus on the limits of small and large~$k_{\perp}\rho_i$. In the former case, $\hat{\Gamma}_0 = k_\perp^2 \rho_i^2/2$. It follows that $\hat{G}=\nabla_\perp^2$, whence \eqref{RMHD1} and \eqref{RMHD2} become the equations of reduced MHD (RMHD)~\cite{Strauss76}. We refer to this limit as the MHD case in the main text. In the opposite limit $k_\perp \rho_i \gg 1$ (the ``kinetic'' case in the main text), $\Gamma_0$ is exponentially small and $\hat{G}= -2/\rho_i^2$ becomes a multiplication by a constant.

We make the ansatz 
\begin{equation}
    \Phi(\psi, \alpha, \ell, t) = \bar{\Phi}(\ell, t)e^{in\alpha},\,\, \Psi(\psi, \alpha, \ell, t) = \bar{\Psi}(\ell, t)e^{in\alpha}\label{eikonal1}
\end{equation}for the solution of \eqref{RMHD1} and \eqref{RMHD2}, where $n\gg 1$ enforces narrow localization in the $\bnabla \alpha$ direction. Eq.~\eqref{eikonal1} fixes the polarization of the wave to be in the $\bb_0\times \bnabla \alpha$ direction; it can be changed by redefining~$\alpha$ to incorporate a shift by a function of~$\psi$ (which leaves $\bB_0$ unchanged).

 The cross product in~\eqref{PoissonBracket1} vanishes after substitution of~\eqref{eikonal1} because perpendicular gradients of $\Phi$ and $\Psi$ are both along $\bnabla \alpha$. Thus, any solution in the form~\eqref{eikonal1} is a nonlinear solution of~\eqref{RMHD1} and~\eqref{RMHD2}. Substituting~\eqref{eikonal1} into~\eqref{RMHD1} and~\eqref{RMHD2}, and taking $\bb_0\times\bnabla_{\perp}$ of the result yields
\begin{align}
    G\left(\frac{n^2|\bnabla\alpha|^2\rho_i^2}{2}\right) \frac{\p u_{\perp}}{\p t} = -n^2 v_A^2\frac{\p}{\p \ell}\left(|\bnabla \alpha|^2 \delta B_{\perp}\right),\label{ueqn} \\ 
    \frac{\p \delta B_{\perp}}{\p t} + \eta n^2|\bnabla \alpha|^2 \delta B_{\perp}=\frac{1}{B_0}\frac{\p (B_0 u_{\perp})}{\p \ell},\label{Beqn}
\end{align}where we have defined $u_{\perp}$ and $\delta B_{\perp}$ such that
\begin{align}
   \delta \bu_{\perp} & = u_{\perp}(\psi, \alpha, \ell, t) e^{in\alpha} \bb_0 \times \bnabla \alpha, \\ \delta \bB_{\perp} & = \delta B_{\perp}(\psi, \alpha, \ell, t) B_0e^{in\alpha} \bb_0 \times \bnabla \alpha,
\end{align}and $G$ is the function satisfying $\hat{G}\Phi = G\Phi$, given by
\begin{align}
    G&\left(\frac{n^2|\bnabla\alpha|^2\rho_i^2}{2}\right) = \frac{2}{\rho_i^2}\left[\Gamma_0\left(\frac{n^2|\bnabla\alpha|^2\rho_i^2}{2}\right)-1\right] \nonumber \\& \to \begin{cases}
			-n^2 |\bnabla \alpha|^2 & \text{as $n^2|\bnabla\alpha|^2\rho_i^2/2\to 0$},\\
            -2/\rho_i^2 & \text{as $n^2|\bnabla\alpha|^2\rho_i^2/2 \to \infty$}.
		 \end{cases}
\end{align}Eqs.~\eqref{ueqn} and~\eqref{Beqn} are identical to Eqs.~\eqref{vorticity} and~\eqref{induction} after normalizing all variables as described in the main text and appending to them the forcing term in Eq.~\eqref{vorticity}.

\vspace{-3mm}
\subsection{2. Evolution of $\Delta \psi$ at $\epsilon \tilde{t}\sim 1$.}
\vspace{-2mm}

For $\epsilon \tilde{t}\sim 1$ (and $\tilde{t}\ll \mathcal{S}$), Eq.~\eqref{usolninnerMHD} has an additional term involving the time derivative of $\tilde{F}$. Including this term, Eq.~\eqref{matchingcond} becomes
\begin{equation}
    \lim_{\tilde{\ell} \to 0} \tilde{u}_{\mathrm{out}} = - \frac{\mathcal{J}_0}{|\Delta '|} 2\epsilon^2 \tilde{t} e^{-\epsilon^2 \tilde{t}^2}+  \frac{\mathcal{S}}{|\Delta '|}\frac{\dd}{\dd\tilde{t}}\lim_{\tilde{\ell}\to 0}\frac{\p \tilde{u}_{\mathrm{out}}}{\p \tilde{\ell}}.\label{early_matching_condition}
\end{equation}We now describe how the inclusion of this term changes evolution of $\Delta \psi$, which, from substituting Eq.~\eqref{JsolninnerMHD} into Eq.~\eqref{Deltapsi}, is given by
\begin{equation}
    \Delta \psi = \frac{1}{|\Delta'|}\left[J_{\parallel 0}(\tilde{t})-\mathcal{J}_0 (1-e^{-\epsilon^2 \tilde{t}^2})\right],\label{Deltapsiearly}
\end{equation}where we have taken $\tilde{F} = (1-e^{-\epsilon^2 \tilde{t}^2})\tilde{f}(\ell)$, for some function $\tilde{f}(\ell)$ (chosen to be $\propto \tilde{l}^3 e^{-\ell^2}$ in the plots shown in the main text).

\textit{MHD case:}\quad Substituting~\eqref{MHD_limx0_u} into \eqref{early_matching_condition}, we find that ${p_n = n}$ for integers $n\geq 0$, and the $C_n$ are determined by
\begin{equation}
    C_{2n+3} = \frac{C_{2n+1}}{\tau_{\mathrm{rec}}^2} + g_n \frac{\mathcal{J}_0 \epsilon^{2+2n}}{\tau_{\mathrm{rec}} \mathcal{S}}, \quad C_{2n+1} = -\frac{C_{2n}}{\tau_{\mathrm{rec}}},\label{MHDearlyrescursion}
\end{equation}where $g_n = (-1)^n \Gamma(2+2n) /\Gamma(1+n)$. Solving Eq.~\eqref{MHDearlyrescursion} for the first few $C_n$ and substituting ${J_{\parallel 0}(\tilde{t}) = -2\mathcal{S}\sum_n  C_n \tilde{t}^{p_n}/ \Gamma (1+p_n)}$ into Eq.~\eqref{Deltapsiearly}, we find
\begin{equation}
    \Delta \psi = -\frac{\mathcal{J}_0}{|\Delta '|} \left[\frac{\epsilon ^2 \tilde{t}^3}{3\tau_{\mathrm{rec}}} - \frac{\epsilon^2 \tilde{t}^4}{12 \tau_{\mathrm{rec}}^2}-\frac{\epsilon^2 \tilde{t}^5}{60\tau_{\mathrm{rec}}^3}( 6\tau_{\mathrm{rec}}^2 \epsilon^2 -1)\right] + O(\tilde{t}^6).\label{deltapsiearlyMHD}
\end{equation}Thus, as $\tilde{t}\to 0$, $\Delta \psi \propto \epsilon ^2 \tilde{t}^3/\tau_{\mathrm{rec}}$, which is the scaling observed in Fig.~\ref{fig:Deltapsi}. This persists until either $\tilde{t}\sim \tau_{\mathrm{rec}}$ (at which point the second term in~\eqref{deltapsiearlyMHD} becomes important) or $\epsilon \tilde{t} \sim 1$ (when the third term does).

\textit{Kinetic case:}\quad Substituting~\eqref{kinetic_limx0_u} into \eqref{early_matching_condition}, we find that $p_n = n/2$ for integers $n\geq 0$, and the coefficients $C_n$ are determined by the relations
\begin{equation}
    C_{4n+5} = -\frac{C_{4n+2}}{\tau_{\mathrm{rec}}^{3/2}} - \frac{2 g_n}{\sqrt{\pi}} \frac{\mathcal{J}_0 2\epsilon^{2+2n}}{\tau_{\mathrm{rec}}^{3/2}|
\Delta '| }, \quad C_{m+3} = -\frac{C_{m}}{\tau_{\mathrm{rec}}^{3/2}}\label{kineticearlyrecursion}
\end{equation}for all $n$ and all $m \neq 2+4n$. Solving Eq.~\eqref{kineticearlyrecursion} for the first few $C_n$ and substituting into Eq.~\eqref{Deltapsiearly} $ {J_{\parallel 0}(\tilde{t}) = -\sqrt{\pi \mathcal{S}/\tilde{t}}\sum_n  C_n \tilde{t}^{p_n}/ \Gamma (1/2+p_n)}$, we find
\begin{equation}
    \Delta \psi = -\frac{\mathcal{J}_0 \epsilon^2}{|\Delta '| \tau_{\mathrm{rec}}^{3/2}}\left[\frac{2 \tilde{t}^{7/2}}{\Gamma(9/2)} - \frac{\tilde{t}^5}{60\tau_{\mathrm{rec}}^{3/2}} - \frac{12 \epsilon^2 \tilde{t}^{11/2}}{\Gamma(13/2)}\right] + O(\tilde{t}^6).\label{deltapsiearlykinetic}
\end{equation}As $\tilde{t}\to 0$, $\Delta \psi \propto \epsilon ^2 \tilde{t}^{7/2}/\tau_{\mathrm{rec}}^{3/2}$, as in Fig.~\ref{fig:Deltapsi}. As in the MHD case~\eqref{deltapsiearlyMHD}, this scaling persists until either $\tilde{t}\sim \tau_{\mathrm{rec}}$ or $\epsilon \tilde{t} \sim 1$.

\bibliography{decay_mod_prx.bib}

\end{document}